\documentclass{aa}
\usepackage{amssymb}
\usepackage{amsmath}
\usepackage{xspace}
\usepackage{graphicx}
\usepackage{enumerate}   
\usepackage{color,xcolor}
\usepackage{txfonts}

\usepackage{natbib}
\usepackage[colorlinks=true,citecolor=blue,urlcolor=blue,linkcolor=blue]{hyperref}

\def\gr{$\gamma$-ray}
\def\GR{$\gamma$ ray}
\def\grb{GRB~221009A}

\begin{document}

\title{Constraint on intergalactic magnetic field from Fermi/LAT observations of the ``pair echo'' of GRB~221009A}

\author{Ie.Vovk$^{1}$ 
, A.Korochkin$^{2}$
, A.Neronov$^{3,4}$
, D.Semikoz$^{3}$
}
\institute{
    Institute for Cosmic Ray Research, The University of Tokyo, 5-1-5 Kashiwa-no-Ha, Kashiwa City, Chiba, 277-8582, Japan
    \and Université Libre de Bruxelles, CP225 Boulevard du Triomphe, 1050 Brussels, Belgium
    \and Université de Paris Cité, CNRS, Astroparticule et Cosmologie, F-75013 Paris, France
    \and Laboratory of Astrophysics, Ecole Polytechnique Federale de Lausanne, CH-1015, Lausanne, Switzerland
}

\authorrunning{Ie.Vovk et al}
\titlerunning{IGMF constaints with GRB221009A}
\abstract
{
Delayed ``pair echo'' signal from interactions of very-high-energy \GR s in the intergalactic medium can be used for detection of the inter-galactic magnetic field (IGMF). We use the data of Fermi/LAT telescope coupled with LHAASO observatory measurements to confirm the presence of IGMF along the line of sight to the \gr\ burst \grb. Comparing the Fermi/LAT measurements with the expected level of the pair echo flux, set by the multi-TeV LHAASO detection, we derive a lower bound $10^{-19}$~G on the IGMF with correlation length $l$ larger than 1 Mpc, improving as $l^{-1/2}$ for shorter correlation lengths. This provides an independent verification of existence of a lower bound on IGMF in the voids of the Large Scale Structure, previously derived from the observations of active galactic nuclei.
}

\maketitle

\section{Introduction}

Very-high-energy (VHE; $\gtrsim 100$~GeV) \GR s, propagating from cosmological distances, suffer from absorption in interactions with infrared and optical photons of extragalactic background light (EBL)~\citep{blumenthal_gould_70, lee98}. This leads to injection of the electron-positron pairs in the intergalactic space and formation of electromagnetic cascades that release the absorbed power as lower energy \gr\ emission as the pairs up-scatter cosmic microwave background photons via Inverse Compton mechanism~\citep{aharonian_coppi}. Development of these cascades is sensitive to the intergalactic magnetic field (IGMF), providing means to measure its properties if a spatially extended and / or time-delayed \gr\ ``pair echo'' is detected from distant VHE sources~\citep{ plaga95, Neronov_2007, neronov_semikoz_09}. 

Lower bound on IGMF has been previously derived from the searches of the extended IGMF-dependent emission around blazars, Active Galactic Nuclei (AGN) with jets aligned along the line of sight~\citep{neronov10, taylor11, tavecchio11, vovk12, fermi_igmf, HEGRA_IGMF, MAGIC_extended, HESS_IGMF, Veritas_IGMF}. These limit suffer from uncertainties of the intrinsic properties of blazars in the multi-TeV energy range: the time-average spectrum and activity duty cycle. 

These uncertainties are reduced in the searches of the delayed IGMF-dependent emission from blazars \citep{2011ApJ...733L..21D,taylor11,fermi_igmf}. Recent analysis of MAGIC telescope data on the blazar 1ES 0229+200 provides a conservative lower bound on the IGMF from the search of the time-delayed signal at the level of $10^{-17}$~G for the long correlation length IGMF \citep{MAGIC:2022piy}. This limit still relies on partial information on the time-average spectrum of the source on decade-long time span (it is not possible to continuously monitor the source in the TeV band on such time scale). 

In this respect, gamma-ray bursts (GRB) may provide a better source type for IGMF searches \citep{razzaque04, ichiki08, murase08, takahashi08, murase09}. Their TeV energy range signal is detectable only during a limited time interval \citep{magic_grb190114c_a, magic_grb190114c_b, hess_grb180720b, hess_grb190829a, magic_grb201216c_gcn}, so that the evolution of the intrinsic source spectrum can be monitored in sufficient details. This information can be used for precision calculation of the expected time-delayed IGMF-dependent flux, to be compared to the observational data~\citep{vovk23}.

In this paper, we apply this idea to the recent exceptionally bright \grb\ that has been detected in the TeV band by LHAASO \citep{2022GCN.32677....1H}. We use publicly available data of Fermi Large Area Telescope~\citep[LAT,][]{2009ApJ...697.1071A}  to search for the pair echo signal from this GRB. Comparing the GRB afterglow signal detected by Fermi/LAT with model predictions of the time-delayed emission from different IGMF parameters, we derive a lower bound on the IGMF strength at the level $10^{-19}$~G. This limit is weaker than the limit from the AGN observations \citep{MAGIC:2022piy}, but provides an independent verification of existence of IGMF in the voids of the Large Scale Structure, obtained with a different type of source with smaller uncertainties on the primary source flux.

\section{Data analysis}

We use publicly available Fermi/LAT data on \grb\ corresponding to {\tt P8R3 SOURCE} \gr\ event selection and collected between 2022-10-09 13:20:39 and 2022-10-30 03:02:45 UTC within $20^\circ$ from the GRB position in the energy range from 100~MeV to 1~TeV. Throughout the analysis we retain events corresponding to the \verb|(DATA_QUAL>0) && (LAT_CONFIG==1)| good-time intervals and the maximal telescope zenith angle of $90^\circ$. Data reduction was performed with Fermitools package v2.0.8 and FermiPy framework\footnote{\url{https://fermi.gsfc.nasa.gov/ssc/data/analysis/}} v1.0.1~\citep{fermipy}, as described in the FermiPy documentation\footnote{\url{https://fermipy.readthedocs.io/}}. We accounted for the galactic ({\tt gll\_iem\_v07.fits}) and extragalactic ({\tt iso\_P8R3\_SOURCE\_V2\_v1.txt}) diffuse emission and included the sources listed the Fermi/LAT fourth source catalogue~\citep[4FGL,][]{4FGL}. The spectral shapes of these sources were taken from the 4FGL~catalogue with only their normalisation left free during the fit. Finally, \grb\ spectrum was assumed to follow the power law form.

\section{Pair echo modelling}

The temporal dependence of the the pair echo signal is set by angular scatter of the cascade electron-positron pairs, caused by both IGMF and angular spreads of the pair production and inverse Compton emission, that are intrinsic to the cascade. This latter alone may result in time delays comparable to the GRB duration~\citep{takahashi08, neronov_semikoz_09}; however for numerical reasons it may be challenging to account for using general-purpose Monte Carlo codes~\citep{vovk23}. We thus use a mixed approach to calculate the expected ``echo'' light curve. 

First we use~\citet{vovk23} approach to model the intrinsically time-delayed emission from electron-positron pairs deposited in the intergalactic medium by interactions of the primary \gr s from the GRB -- without contribution from IGMF. In this model we assume that the intrinsic source spectrum in the TeV range is a power law with the slope and normalization found in the LHAASO data \citep{lhaaso_grb}. The total model flux over the five time bins selected in \citep{lhaaso_grb} was computed. Though no high-energy cut-off was found in the LHAASO data, we conservatively limit the maximal intrinsic photon energy to $E_{max}=10$~TeV; given the soft GRB spectrum in the multi-TeV energy range, this artificially reduces the time-delayed emission (from energies above 10~TeV) by a up to a factor of 2.

Then we run the publicly available CRPropa code~\citep{AlvesBatista:2021mne} to model the time-delayed signal expected in presence of IGMF (omitting the intrinsic cascade scatter). This code has been cross-validated with the other Monte-Carlo modelling codes \citep{Kalashev:2022cja} and is known to provide sufficient precision for sources at moderate redshifts, which is the case for \grb\ at the redshift $z=0.151$ \citep{2022GCN.32648....1D}. 
For the intergalactic magnetic field, we explore different configurations characterised by strength $B$ and correlation length $l$ in the voids of the Large Scale Structure. We model this field as cell-like, with the field being constant within each ``cell'', but randomly changing the orientation from one cell to another.

Finally, the temporal profile of the ``echo'' emission is modelled as a convolution of the intrinsic and IGMF-induced profiles obtained above. The resulting model of the delayed signal is shown in Fig.~\ref{fig:igmf_scan}.

\section{IGMF limit}

\begin{figure}
    \centering
    \includegraphics[width=\linewidth]{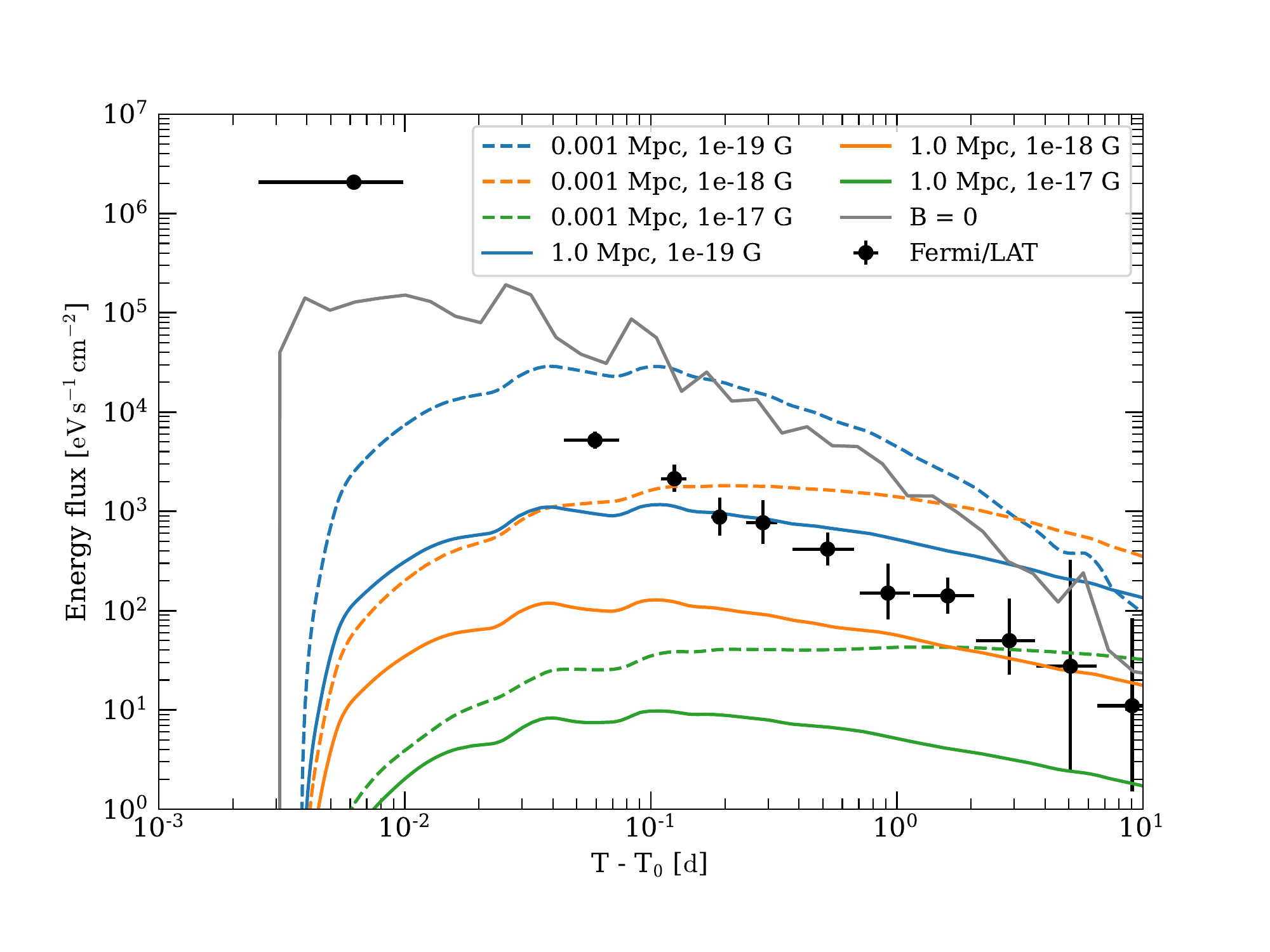}
    \caption{Comparison of the measured \grb\ Fermi/LAT signal in the $0.1-10$~GeV energy range with the models of the pair echo emission for different IGMF correlation length and strength combinations, stemming from the LHAASO measurements of its multi-TeV flux. Short time scale fluctuations visible in the curves are numerical artifacts of the ``echo`` signal calculation.}
    \label{fig:igmf_scan}
\end{figure}

The expected ``echo'' signal for weak (or even zero) IGMF substantially exceeds the Fermi/LAT measurements, as is evident from Fig.~\ref{fig:igmf_scan}.  This rules out the possibility of zero IGMF along the line of sight toward the GRB. 

At the same time, all the models with IGMF strength below $10^{-19}$~G are also inconsistent with the Fermi/LAT measurements of \grb\ flux. The possibility of $B\simeq 10^{-19}$~G is marginally inconsistent with the data in the case of the correlation length $l=1$~Mpc. The expected secondary emission flux is still well above the observed flux limits in the case $l=1$~kpc. This is explained by the fact that shorter correlation length field is less efficient in deflections of electrons and positrons. The secondary flux in the 0.1-10 GeV range is produced via inverse Compton scattering of the Cosmic Microwave Background photons by electrons with energies in the $E_e\simeq 0.3-3$~TeV range \citep{neronov_semikoz_09}. Such electrons loose energy on the inverse Compton scattering on the distance scale of about $D_e\simeq 0.1-1$~Mpc \citep{neronov_semikoz_09}. If the magnetic field correlation length is longer than $D_e$, electrons and positrons are deflected by an angle $\alpha=D_e/R_L$ where $R_L=E_e/eB$ is the giro-radius in magnetic field $B$ and $e$ is the electron charge. The deflection angle does not depend on the magnetic field correlation length. To the contrary, if $l\ll D_e$, the deflection direction changes $n=D_e/l$ times on the cooling distance scales, so that the overall deflection pattern is a random walk in the pitch angle, with the resulting deflection angle $\alpha=(l/R_L)\sqrt{n}\propto B\sqrt{l}$. Stronger magnetic field is required to sufficiently deflect the electrons and reduce the pair echo signal.


\begin{figure}
    \centering
    \includegraphics[width=\linewidth]{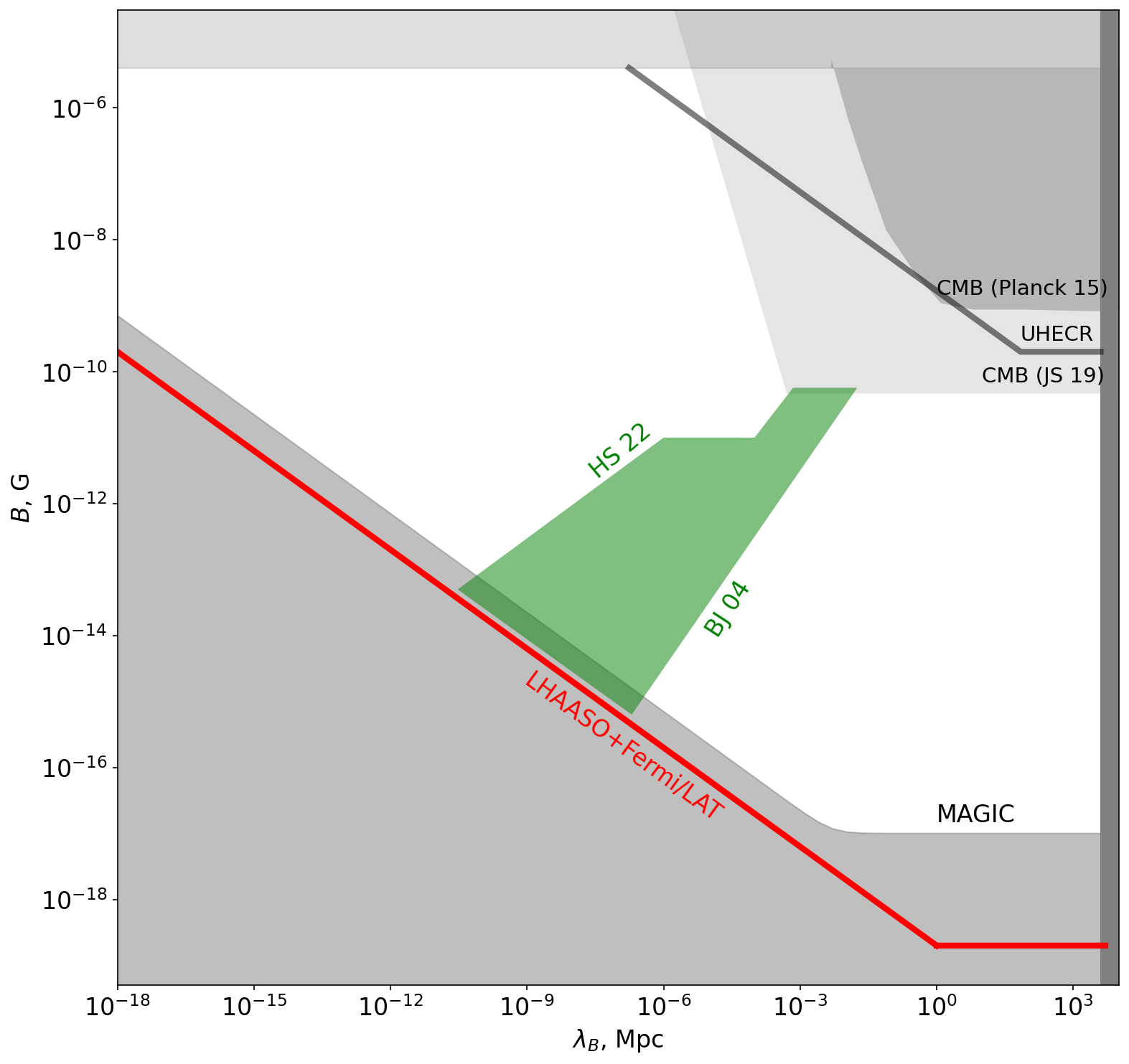}
    \caption{Lower bound on IGMF derived from the \grb\ (red line), compared to existing bounds form \gr, radio, CMB and UHECR observations and predictions of the cosmological evolution models. The CMB upper bounds are from Planck \citep{2016A&A...594A..19P} and from the analysis of \citet{2019PhRvL.123b1301J}. UHECR upper bound is from \citet{2021arXiv211208202N}. MAGIC lower bound is from \citet{MAGIC:2022piy}. Green-shaded area shows the range of predictions for the endpoints of cosmological evolution of primordial magnetic fields. BJ04 is from \citet{2004PhRvD..70l3003B}, HS22 is from \citet{2022arXiv220303573H}.  }
    \label{fig:exclusion}
\end{figure}

Overall, we find that the lower bound on the IGMF is $l$-independent for $l>1$~Mpc and scales as $B>10^{-19}(l/1\mbox{ Mpc})^{-1/2}$~G for shorter correlation lengths, as shown in Fig. \ref{fig:exclusion}. 

\section{Discussion}

The lower bound on IGMF derived from a combination of LHAASO and Fermi/LAT data on \grb, at $B\simeq 10^{-19}$~G level, is weaker than that derived from MAGIC observations of the blazar 1ES~0229+200. The difference between the two bounds is somewhat smaller at shorter correlation lengths, because of the different correlation length dependence of the bounds. The two sources, \grb\ and 1ES~0229+200, are at comparable redshifts. The difference in the correlation length dependence of the lower bound on IGMF is explained by the different energy range in which the time-delayed secondary emission was searched for the two sources. MAGIC search of the pair echo was concentrated on the 100~GeV range, while Fermi/LAT analysis reported in our paper was for the 1~GeV energy range. The cooling distance of electrons generating the inverse Compton emission at 1~GeV is larger so that they can accumulate random deflection by IGMF on longer distance scales.

The \gr\ observations constrain the magnetic field in the voids of the Large Scale Structure. Recent modelling of magnetised outflows from galaxies suggests that the strength of these outflows is not sufficient for filling the voids, so that the void field is most probably of cosmological origin \citep{Marinacci:2017wew}. The GRB 221009A Fermi/LAT pair echo limit on the short correlation length magnetic fields provides a limit on such cosmological IGMF (green-shaded region in Fig. \ref{fig:exclusion}) that is of the same order of magnitude as the AGN limit. 

The cosmological magnetic fields may be produced during the epochs of Electroweak and Quantum Chromodynamics (QCD) transitions in the Early Universe or during the period of Inflation (see  \citet{2001PhR...348..163G,durrer13} for reviews). Maximal initial correlation length of these fields does not exceed the size of the cosmological horizon at the moment of the field generation. This limits the comoving correlation length of the fields from the Electroweak epoch to about $10^2$ astronomical units and the QCD epoch field to about a parsec. Turbulent decay of the field from the moment of generation up to the recombination epoch leads to decrease of the field strength and increase of the correlation length up to the scale of the largest processed eddies, $l\simeq 1(B/10^{-8}\mbox{ G})$~Mpc \citep{2004PhRvD..70l3003B} (the boundary of the green-shaded region marked BJ04). If the turbulent decay is governed by the magnetic reconnection, somewhat shorter final correlation length is expected \citep{2022arXiv220303573H} (the boundary marked HS22). The GRB pair echo \gr\ data limit the parameters of the cosmological magnetic field to be  $B>10^{-15}\mbox{ G}$, $\lambda_B>0.1\mbox{ pc}$ for the \citet{2004PhRvD..70l3003B} evolution model and  $B>10^{-14}\mbox{ G}$, $\lambda_B>10^{-5}\mbox{ pc}$ for the \citet{2022arXiv220303573H} evolution model.

When this study was at final stage waiting for LHAASO data, two other publications arrived using Elmag code \citet{Dzhatdoev:2023opo, Huang:2023uhw}. Note that results of those publications are not completely consistent between each other giving limits between $B>3\times10^{-19}$ G and $B>10^{-18}$ G. In comparison, our result benefits from a cross-check between CRpropa and CRbeam codes and, importantly, accounts for the intrinsic time delay, substantially diluting the ``echo'' flux on time scales below a day and thus yielding more conservative limit $B>10^{-19}$ G.

\begin{acknowledgements}
The work of D.S. and A.N. has been supported in part by the French National Research Agency (ANR) grant ANR-19-CE31-0020. IV gratefully acknowledges the support of the CTA-North computing center at La Palma, Spain for providing the necessary computational resources.
\end{acknowledgements}

\bibliographystyle{aa}
\bibliography{references.bib}

\end{document}